\documentstyle[12pt]{article}
\advance\voffset -2.0 truecm       
\title{\bf AGB Stars: What Should Be Done~?}
\author{C.~A.~Frost \& J.~C.~Lattanzio\\
\vspace{1cm}\\
\normalsize Department of Mathematics, Monash University, Australia\\
\normalsize and\\
\normalsize Institute of Astronomy, Cambridge University\\}
\date{\mbox{}}
\begin{document}
\maketitle
\pagestyle{empty}
%
%
\input epsf
%
%
\def\bull{\vrule height .9ex width .8ex depth -.1ex}
\makeatletter
\def\ps@plain{\let\@mkboth\gobbletwo
\def\@oddhead{}\def\@oddfoot{\hfil\tiny\bull\quad
``Stellar Evolution : What Should Be Done'';
32$^{\mbox{\rm nd}}$ Li\`ege\ Int.\ Astroph.\ Coll., 1995\quad\bull}%
\def\@evenhead{}\let\@evenfoot\@oddfoot}
\makeatother
%
%
\def\beginrefer{\section*{References}%
\begin{quotation}\mbox{}\par}
\def\refer#1\par{{\setlength{\parindent}{-\leftmargin}\indent#1\par}}
\def\endrefer{\end{quotation}}
%
%
{\noindent\small{\bf Abstract:} 
We provide an overview of the current theoretical picture of AGB stars,
with particular emphasis on the nucleosynthesis occurring in these stars, both
in their deep interiors, associated with thermal pulses or flashes, and also 
during the phase of ``hot bottom burning''. 
These processes are illustrated with some 
new results from hot bottom burning calculations. Finally,
we conclude with recommendations about
``what should be done''.
%
%
%
\def\msun{M_\odot}
\def\lsun{L_\odot}
\def\H1{$^1$H}
\def\Hethree{$^3$He}
\def\He4{$^4$He}
\def\Li7{$^7$Li}
\def\Be7{$^7$Be}
\def\C12{$^{12}$C}
\def\Cthirteen{$^{13}$C}
\def\N14{$^{14}$N}
\def\Nfifteen{$^{15}$N}
\def\O16{$^{16}$O}
\def\Oseventeen{$^{17}$O}
\def\Oeighteen{$^{18}$O}
\def\Ne22{$^{22}$Ne}
\def\F19{$^{19}$F}
\def\Mg24{$^{24}$Mg}
\def\Al26{$^{26}$Al}
\def\Si30{$^{30}$Si}
\def\Mbol{M$_{bol}$}
%
%
\section{Introduction}
In the last twenty years much research has been dedicated to
the understanding of Asymptotic Giant Branch (AGB) stars. For 
particularly noteworthy reviews see Iben and Renzini (1983),
Lattanzio (1989), Sackmann and Boothroyd (1991) and Iben (1991).
Recent evidence for extensive nucleosynthesis at the bottom of their
deep convective envelopes (known as ``Hot Bottom Burning'', HBB)
together with extensive data from isotopic analysis of grains
in meteorites is leading to a revolution in 
the quantitative demands being placed on the models. Further,
the discovery that the \Cthirteen~pocket is burned
under radiative conditions rather than in the intershell convective
zone (see below for details) demands that we re-examine the
models and their nucleosynthesis.

We  begin by giving a qualitative analysis of the evolution of
stars of masses $\simeq 1 \msun$ and $\simeq 5\msun$ in Section~2.
This illustrates
the cases where we do, and do not, find the ``second dredge-up'',
and introduces the basic principles of the evolution of all stars which 
spend some time on the AGB. These vary from a minimum mass 
(probably a little under $1 \msun$)
to a maximum mass of $M_{up}$, which just avoids
core carbon ignition, and is about $9 \msun$
(depending on composition). 

Section 3 will  outline the basic evolution
during a thermal pulse, but quite briefly because 
this is well understood (or, rather, as well 
understood as it is likely to be for the present!). As an
illustration of the nucleosynthesis which can occur during this stage, we will
explicitly discuss the dredge-up of \C12~and the formation of Carbon stars.
Section 4 will discuss the $s$-process, why we believe that \Cthirteen~is
the neutron source, and how we believe the \Cthirteen~is produced. Here
we will also discuss the problem of \F19~production.
In Section 5 we will explain the observational motivation for considering
HBB, and its consequences for the composition of the star. 
Particular emphasis will be placed on \Li7~production and how HBB can
prevent the formation of Carbon stars.
Section 6 will introduce meteorite grains 
as an important source of abundance information, which is driving
models to a higher level of precision.
Finally, in Section 7, we will
discuss the immediate future:~``what should be done?''.

\section{Basic Stellar Evolution}
We now give a qualitative overview of the evolution
of stars of masses $1 \msun$ and $5 \msun$, with emphasis
on the mechanisms and phenomenology of the structural and evolutionary
changes. In this section we consider only the evolution up to the 
beginning of thermal pulses on the AGB (the ``TP-AGB'').

\subsection{Basic Evolution at $1 \msun$}
We make the usual assumption that a star reaches the zero-age
main sequence
with a homogeneous chemical composition (for an alternative evolutionary
scenario see Lattanzio 1984). Figure~1 shows a schematic HR diagram
for this star. Core H-burning occurs radiatively, and the central
temperature
and density grow in response to the increasing molecular
weight (points 1--3).
At central H exhaustion
(point 4)
the H profile is as shown in inset~(a) in Figure~1. The star now
leaves the main sequence and crosses the Hertzsprung Gap
(points 5--7), while the central
\He4 core becomes electron degenerate and the nuclear burning is established
in a shell surrounding this core. Inset~(b) shows the advance of the H-shell
during this evolution. Simultaneously, the star
is expanding and the outer layers become convective. As the star reaches
the Hayashi limit ($\sim$ point 7), convection extends quite deeply
inward (in mass) from the surface, and the star
ascends the (first) giant branch. The
convective envelope penetrates into the region where partial H-burning has
occurred earlier in the  evolution, as shown in inset~(c) of Figure~1. This
material is still mostly H, but with added
\He4~together with the products of CN cycling, primarily
\N14~and \Cthirteen. These are now mixed to the surface (point~8)
and this phase
is known as
the ``first dredge-up''. The most important surface abundance
changes are an increase in the \He4~mass fraction
by about 0.03 (for masses less than about $4\msun$),
while \N14~increases at the expense of \C12 by around
30\%, and the number ratio \C12/\Cthirteen~drops from its initial
value of $\sim 90$ to lie between
18 and 26 (Charbonnel 1994).

\begin{figure}
\epsfysize = 0.90\vsize
\centerline{\quad\epsffile{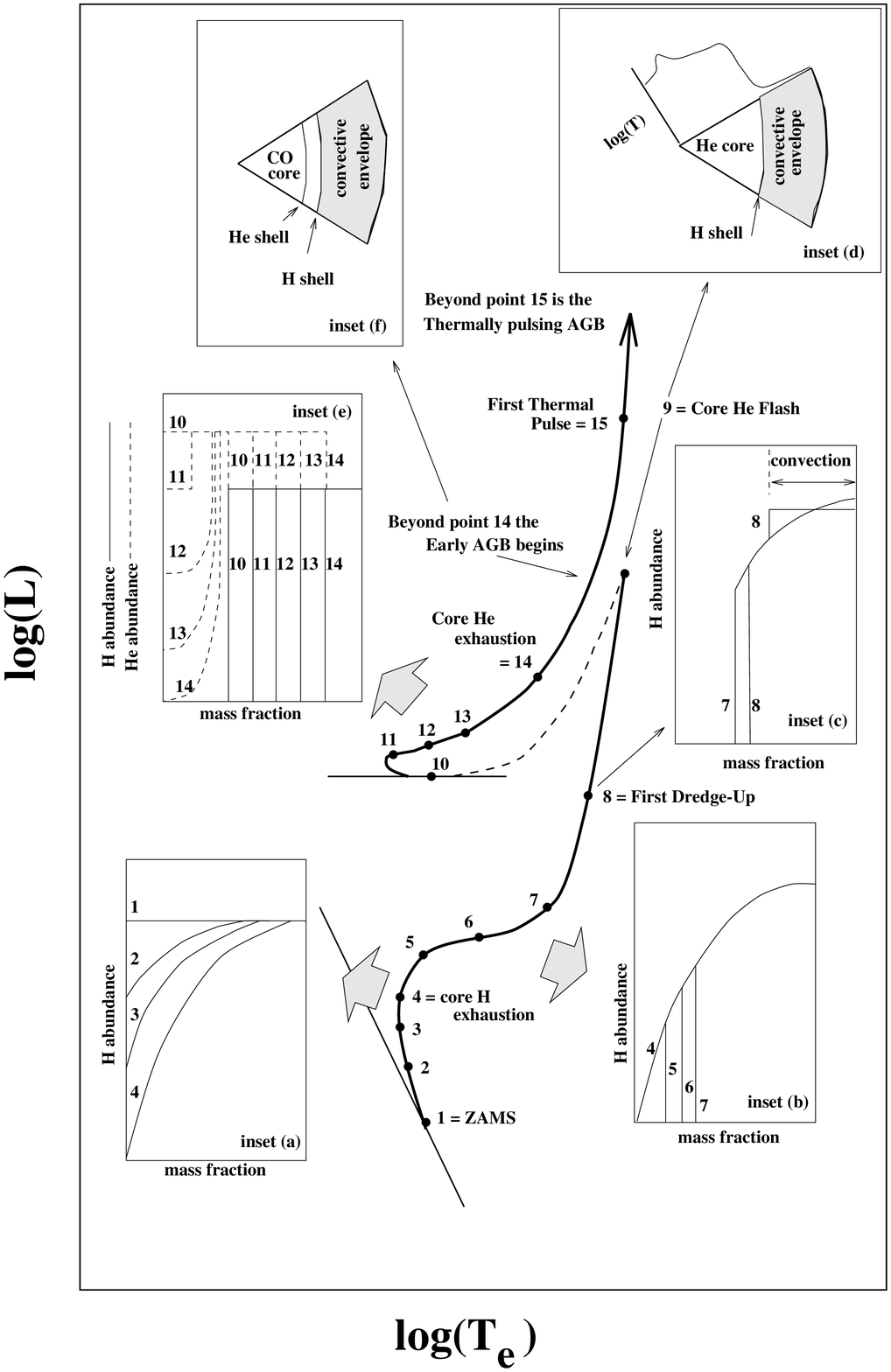}}
\smallskip
\caption{Schematic of evolution at $\sim 1\msun$.}
\end{figure}

        As the star ascends the giant branch the \He4-core continues to
contract and heat. Neutrino energy losses
from the centre cause the temperature maximum to move outward, as shown
in inset~(d) of Figure~1.
Eventually triple alpha reactions are ignited at this point
of maximum temperature, but with
a degenerate equation of state. The temperature and density are 
decoupled:~the resulting ignition is violent, and is referred to as the
``core helium flash'' (point~9: see for example Deupree 1984).
Following this the star quickly
moves to the Horizontal Branch where it burns \He4~gently in a convective
core,
and H~in a shell (which provides most of the luminosity).
This corresponds to points 10--13 in Figure~1.
Helium
burning increases the mass fraction of \C12~and
\O16~(the latter through
\C12$(\alpha,\gamma)$\O16) and
the outer regions of the convective core become
stable to the Schwarzschild convection criterion but unstable to that of
Ledoux:~a
situation referred to as ``semiconvection'' (space prohibits a discussion
of this phenomenon, but an excellent physical description is contained in
Castellani {\it et al.} 1971a,b). 
The semiconvection causes the composition profile to adjust itself
to produce convective neutrality,
with the resulting profiles as shown in inset~(e) of Figure~1.\par

        Following \He4 exhaustion (point 14) the star ascends the giant
branch for the second time, and this is known as the Asymptotic Giant
Branch, or AGB, phase. The final proportions of \C12~and
\O16 in the \He4-exhausted core
depend on the uncertain  rate for the \C12$(\alpha,\gamma)$\O16
reaction (see Arnould 1995).
The core now becomes electron
degenerate, and the star's energy output is provided by the
\He4-burning shell (which lies immediately above the C-O core) and
the H-burning shell. Above both is the deep convective envelope.
This structure is shown in inset~(f) in Figure~1.  We will
later see that the
\He4-shell is thermally unstable, as witnessed by the recurring ``thermal
pulses''. Thus the AGB is divided into two regions: the early-AGB, prior to
(and at lower luminosities than)
the first thermal pulse, and the thermally-pulsing AGB (TP-AGB)
beyond this. We will return to this in section 3.\par

\subsection{Basic Evolution at $5 \msun$}

         A more massive star, say of $\sim 5\msun$,
begins its life very similarly
to the lower mass star discussed above. The main initial difference is that
the higher temperature in the core causes CNO cycling to be the main
source of H-burning, and the high temperature dependence of these
reactions causes a convective core to develop. As H is burned into \He4
the opacity (due mainly to electron scattering, and hence proportional
to the H content) decreases and the extent of the convective core
decreases with time. This corresponds to points 1--4 in Figure~2.
Following core
H exhaustion there is a phase of shell burning as the star crosses
the Hertzsprung Gap (points 5--7 and inset~(b)), and then
ascends the (first) giant
branch. Again the
inward penetration of the convective envelope (point 8)
reaches regions where there
has been partial H-burning earlier in the evolution, and thus these
products
(primarily \Cthirteen~and
\N14, produced at the cost of \C12) are mixed to the surface
in the first dredge-up, just as seen at lower masses, and sketched in
inset~(c)
of Figure~2.\par

      For these more massive stars
the ignition of \He4~occurs in the centre and under non-degenerate
conditions, and the star settles down  to a period of quiescent
\He4-burning in a  convective core, together with
H-burning in a shell (see inset~(d) in Figure~2).
The competition between these two energy sources determines the
occurrence and extent of the subsequent blueward excursion in the HR
diagram ({\it e.g.\/} Lauterborn {\it et al.\/} 1971),
when the star crosses the instability strip and is observed as a
Cepheid variable (points 10--14). Following core \He4 exhaustion the
structural re-adjustment
to shell \He4 burning results in a strong expansion, and the H-shell is
extinguished as the star begins its ascent of the AGB. With this entropy
barrier removed, the inner edge of the convective envelope is free
to penetrate the erstwhile H-shell. Thus the products of complete
H-burning are mixed to the surface in what is called the ``second
dredge-up'' (point 15). This again alters the surface compositions of \He4,
\C12, \Cthirteen~and \N14,
and actually reduces the mass of the H-exhausted core, because in the
process of mixing \He4~outward we also mix H~inward (see
inset~(e) in Figure~2). Note that there is a critical mass (of
about 4$\msun$, but dependent on composition) below which the
second dredge-up does not occur.
Following dredge-up the H-shell is re-ignited and
the first thermal pulse
occurs soon after:~the star has reached the thermally-pulsing
AGB, or TP-AGB.
Note that at this stage
the structure is
qualitatively similar for all masses.\par

\begin{figure}
\epsfysize = 0.9\vsize
\centerline{\quad\epsffile{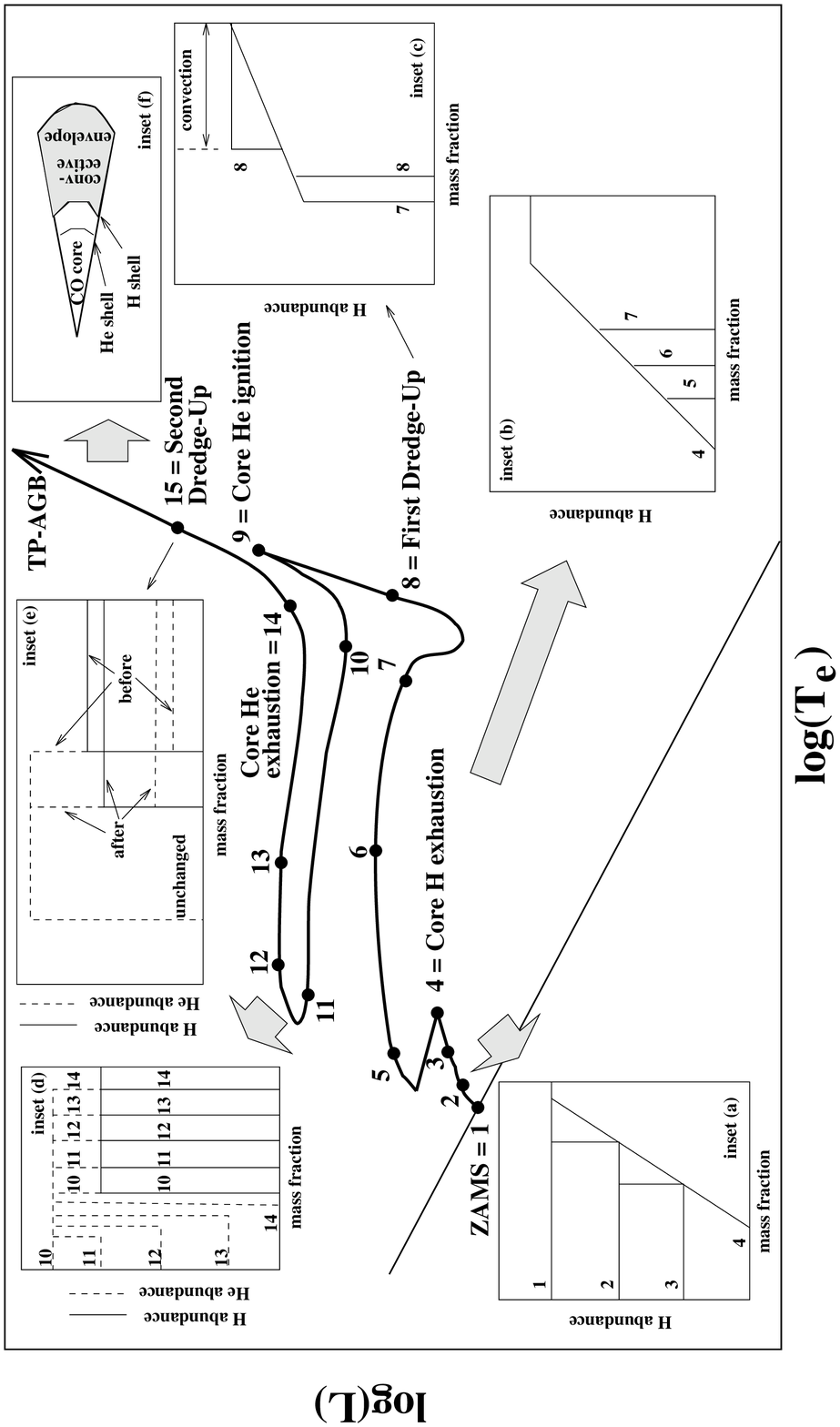}}
\smallskip
\caption{Schematic of  evolution at $\sim 5\msun$.}
\end{figure}

\section{Thermal Pulses on the AGB: Making Carbon Stars}

This phase has been reviewed extensively, and we present here only a very brief
summary (for further details, see Iben and Renzini 1983,
Lattanzio 1989, Sackmann and Boothroyd 1991, and Iben 1991).
The \He4-burning shell
is thermally unstable ({\it e.g.\/} Schwarzschild and H\"arm 1965,
Sackmann 1977, Sugimoto and Fujimoto 1978),
and experiences periodic outbursts called ``shell flashes''
or ``thermal pulses''.
The four phases of such a thermal pulse
are: (a) the off phase, where the
structure is basically that of an early-AGB star. 
During this phase almost all of the surface  luminosity is  provided by  
the H-shell. This phase lasts for $10^4$ to $10^5$ years, depending on the
core-mass; 
(b) the ``on'' phase, 
when the \He4-shell burns very strongly, 
producing luminosities up to $\sim 10^8\lsun$. The
energy deposited by these \He4-burning reactions 
is too much for radiation to carry, and a 
convective shell develops, which extends from the \He4-shell almost
to the H-shell. This convective zone is comprised mostly of \He4~(about 75\%)
and \C12~(about 22\%), and lasts for about 200 years; 
(c) the ``power down'' phase, where the \He4-shell 
begins to die down, and the intershell convection is shut-off.
The previously released energy drives a substantial expansion, pushing the
H-shell to such low temperatures and densities that it is extinguished (or 
very nearly so); and~(d)
the ``dredge-up'' phase, where the convective envelope, in response to the 
cooling of the outer layers, extends inward and, in later pulses, beyond the
H/He interface (which was previously the H-shell) 
and can even penetrate the erstwhile flash-driven convective zone. 
This results in the \C12~which was produced by the \He4-shell, and mixed
outward
by the flash-driven convection, now  being mixed to the surface by the envelope 
convection. This is the ``third dredge-up'', and it
qualitatively (and almost quantitatively) accounts for the
occurrence of Carbon stars at higher luminosities on the AGB.
Figure~3 shows these four phases during one
pulse (top) and during two consecutive pulses (bottom). 
From this figure we see the definition of the so-called ``dredge-up 
parameter'', $\lambda$. This is defined as
$\lambda = {\Delta M_{dredge}}/{\Delta M_H}$ where $\Delta M_{dredge}$ is the
amount of mass dredged-up by the convective envelope, and $\Delta M_H$ is
the amount of mass through which the H-shell has moved during the
interpulse phase. Typical evolutionary calculations show that 
$\lambda \simeq 0.3$ (at least for lower masses, although we find
$\lambda \sim 0.9$ for $M \sim 6\msun$).
Also shown in the
bottom panel is the variation of the total radiated luminosity
and the two nuclear energy 
sources (i.e. the luminosities from H~and \He4~burning) during a
pulse cycle.\par

\section{Interior Nucleosynthesis on the AGB}

\subsection{\Cthirteen~and the $s$-process}

\begin{figure}
\epsfysize = 0.9\vsize
\centerline{\quad\epsffile{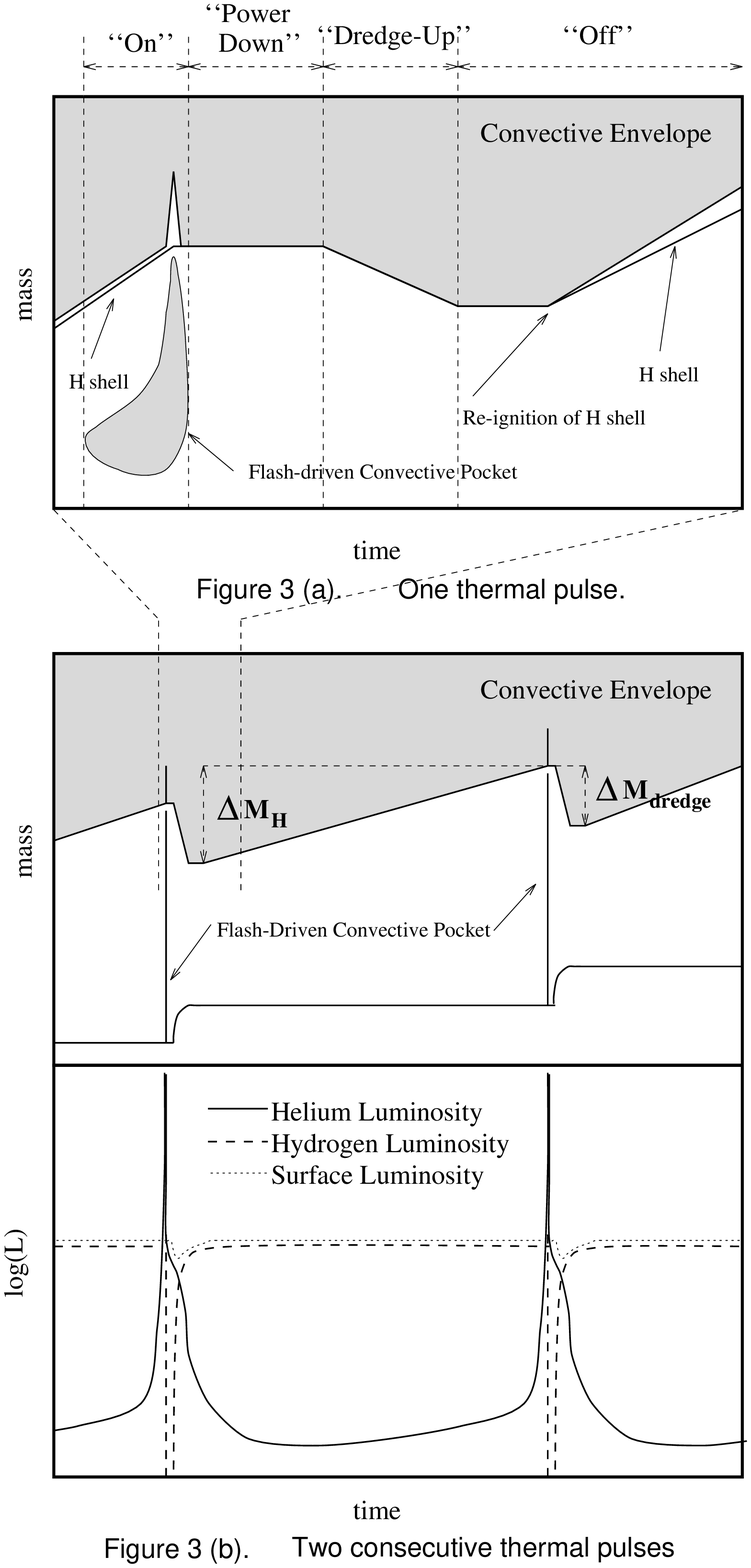}}
\smallskip
\caption{Details of interior evolution during one thermal pulse (top) and
 for two consecutive pulses (bottom).}
\end{figure}

     It is now well established observationally that 
many AGB stars show an overabundance of the
$s$-process elements. This is easily understood within the
picture given above. It was initially envisaged ({\it e.g.\/} Iben 1975a) that the
H-shell, which burns primarily by the CNO cycles, would leave
behind significant amounts of \N14. During the next flash cycle
this \N14~is mixed downward to regions where the temperature is higher
(the exact value depending mainly on the star's core mass) and 
then the sequence of reactions
$$^{14}{\rm N}(\alpha,\gamma)^{18}{\rm F}(\beta^+\nu)^{18}{\rm O}
(\alpha,\gamma)^{22}{\rm Ne}$$
occurs. If the core mass is greater than about $0.9 \msun$ then this is 
followed by \Ne22$(\alpha,{\rm n})^{25}$Mg which releases neutrons
that are then captured by many species, including $^{56}$Fe and its progeny,
producing a distribution of $s$-process elements
which is close to that seen in the
solar system (Iben 1975b, Truran and Iben 1977).\par

Various observations (see Smith {\it et al.\/} 1987)
indicate that the $s$-process enriched stars have masses $\sim 1-3 
\msun$, which means they have smaller cores and consequently cooler intershell
convection zones. Thus the \Ne22~source would
never be activated (or, at least, not at a sufficient rate to
provide enough neutrons for the observed $s$-processing to occur).
Hence it appears that \Ne22~is not the neutron source, and we are forced to 
find another. One rather
obvious source is $^{13}{\rm C}(\alpha, {\rm n})^{16}$O, which
ignites at much lower temperatures. 
The problem here is to produce enough 
\Cthirteen~to provide sufficient neutrons. 
The obvious source is CN cycling in the H-shell, but this leaves behind only 
very small amounts of
\Cthirteen:~$X($\Cthirteen)$\sim 10^{-2} X($CNO).

Sackmann (1980) and Iben (1982) discussed the possibility of post-pulse
expansion causing the carbon rich region to be exposed to very low
temperatures, with a consequent increase in the opacity due to
Carbon recombination, and
possibly leading to some mixing.
Iben and Renzini (1982a) indeed  
showed that following a pulse
the bottom of the convective envelope can 
become semiconvective. This results in
the diffusion of  some protons downward beyond the formal maximum inward
extent of the convective envelope during the third dredge-up phase. This
is shown schematically in Figure~4. The protons which are deposited by this 
semiconvection are in a region comprising about 75\% \He4~and 22\% \C12,
so when the H-shell is re-ignited they are burned into 
\Cthirteen~(and \N14).
In this scenario, which we shall call the ``classical \Cthirteen~scenario'',
when the next thermal pulse occurs
the \Cthirteen~is engulfed by the flash-driven convection, and then
in this \He4-rich environment neutrons are released by the
\Cthirteen$(\alpha,
$n$)$\O16. These neutrons are then captured by $^{56}$Fe and its progeny
to produce the observed $s$-process elements (see Iben and Renzini 1982b,
Gallino {\it et al.\/} 1988).\par

\begin{figure}
\epsfysize = 0.33\vsize
\centerline{\quad\epsffile{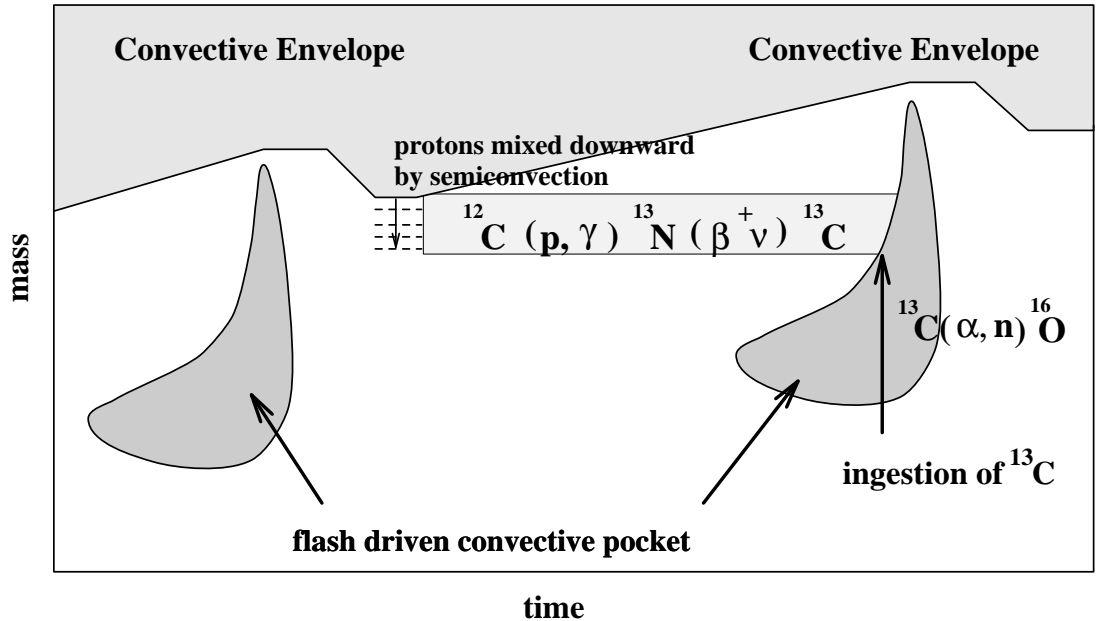}}
\smallskip
\caption{Schematic of the ``classical'' scenario describing how
\Cthirteen~acts as a neutron source in low mass AGB stars.}
\end{figure}

    This scenario has many attractive features, but it has always had some
problems (Lattanzio 1989), the most serious 
of which is that not all calculations
reproduce this semiconvective mixing. Of course, a small amount of
overshoot\footnote{Note that it is incorrect to refer to this as 
``undershoot''. Overshoot refers to the mixing {\it beyond\/} the formal 
convective boundary, and undershoot would mean that the mixing ended
{\it before\/} reaching the normal boundary. Thus the phenomenon of
overshooting in the inward direction is quite distinct from undershoot.}
inwards could produce 
the same results, as could almost any form
of mixing which will distribute some H~below the convective envelope and 
into the
previously flash-driven convective zone. In any event, to calculate the
effects of this proposed mixing, it has been common to
artificially add a \Cthirteen~profile just before a pulse. 
This is how subsequent
nucleosynthesis was calculated in the classical scenario.

A modification to this scenario appears to have been found
by Straniero {\it et al.\/} (1995a,b) who discovered that
any \Cthirteen~present will burn under {\it radiative\/} conditions during the
interpulse phase. They observed this to happen in 
their calculation of a $3 \msun$ model with $Z=0.02$. 
The temperature of the intershell region
is usually lower during the interpulse phase
than when this zone becomes convective during the next pulse. 
But it does increase during the interpulse phase, 
reaching values of $T_6 \sim 90$ just before the later pulses. 
With an interpulse
period of about 50,000 years there is
plenty of time for the \Cthirteen~to be 
consumed by alpha captures between pulses and hence
under radiative conditions.
Thus  all the \Cthirteen~is burned into 
\O16~before
the next pulse, by the same
\Cthirteen$(\alpha,$n$)$\O16~reaction as in the
classical scenario.
But now this occurs at lower temperatures,
and with the release of neutrons {\it in situ\/}, so that the 
neutron density remains
very low, with $n_n$ at most a few $\times 10^7$, compared with
$\sim 10^9$ in the classical picture.
It now appears that an asymptotic distribution of $s$-elements is
achieved after fewer pulses (about 5, see Gallino and Arlandini 1995) 
than in the
classical scenario. The resulting $s$-element distribution looks similar
to that in the classical scenario only for the heavier elements, with 
significant differences appearing for the isotopes with $A < 90$. For
further details refer to Gallino and Arlandini (1995).

It is worth noting, finally,  that the radiative burning of \Cthirteen~has
been confirmed by Mowlavi {\it et al.\/} (1995) and in unpublished calculations
by the authors.

\subsection{The Production of \F19}

     Jorissen {\it et al.\/}~(1992) discovered  that the 
\F19/\O16~ratio in AGB stars
increases with the \C12/\O16~ratio 
implicates thermal pulses
in the origin of this \F19. Little
theoretical work has been done at this stage.
The paper by Jorissen {\it et al.\/} investigated many possible scenarios, and
this was followed by Forestini {\it et al.\/} (1992)  who investigated the most
promising scenario in more detail. This is shown in Figure~5. 
Here, some \Cthirteen~produces neutrons via the \Cthirteen$(\alpha,
$n$)$\O16~reaction discussed above, and some of these neutrons
are captured by \N14~to produce $^{14}$C and protons. These protons,
plus possibly some from \Al26$($n,p$)^{26}$Mg, are then captured by
\Oeighteen~and the sequence
\Oeighteen$($p$,\alpha)$\Nfifteen$(\alpha,\gamma)$\F19~produces 
the observed \F19, which is then dredged to the surface  in the
usual way following the pulse. For all except those stars with the 
highest abundances of \F19~it appears that  the amount of \Cthirteen~left
from the CN cycling H-shell is sufficient. This may be important
in view of the fact that \Cthirteen~is now believed to burn 
between pulses, and may
indicate that small overabundances of \F19~are easily explained without
the extra \Cthirteen~provided by the semiconvection of 
Iben and Renzini (1982a,b).
Yet for those stars showing more enhanced \F19~we may need to invoke some
extra-mixing (semiconvection, overshoot, diffusion, or whatever) to
distribute some H~into the carbon-rich intershell region. 

\begin{figure}
\epsfysize = 0.35\vsize
\centerline{\quad\epsffile{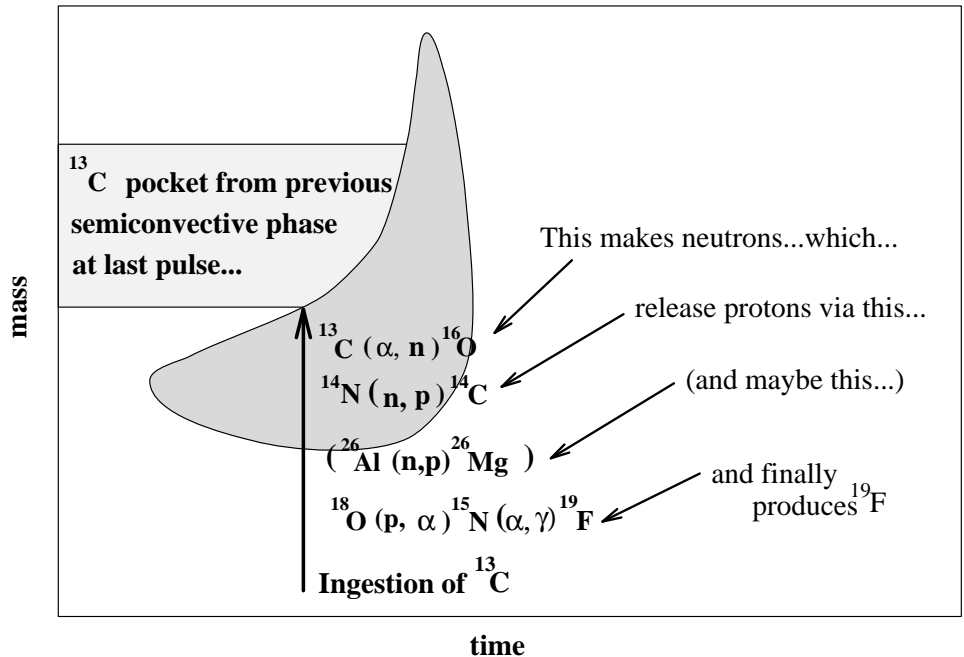}}
\smallskip
\caption{Schematic of the \F19~production mechanism as envisaged 
by Jorissen {\it et al.\/} (1992) and Forestini {\it et al.\/} (1992)}
\end{figure}

Recent models by Mowlavi {\it et al.\/} (1995) show that the scenario described above
can work only for the first few pulses (for low masses). After that the
high temperature in the intershell destroys \F19~via
$^{19}{\rm F}({\rm p},\alpha)^{16}{\rm O}$. 
Furthermore, with the new (and much higher) rate for the $^{18}$O($\alpha,
\gamma)^{22}$Ne reaction, the survival of sufficient \Oeighteen~is not
assured. To complicate matters further, during these thermal
pulses the lifetime of (neutrons and) protons can actually decrease
below the convective timescale, so that the usual homogeneous mixing 
approximation breaks down and one must include some time-dependent mixing
algorithm, such as the diffusion approximation or perhaps something else ({\it e.g.\/} 
Cannon {\it et al.\/} 1995). Clearly we have not yet heard the last word about
\F19 and much work still needs to be done to clarify the
situation.

\subsection{Producing heavier elements}

Stellar evolutionary calculations 
include all nuclear reactions necessary
to calculate the energy production in stellar models. They usually ignore 
the many other reactions which are energetically negligible. However, with
improved observations and the emerging science of isotopic analysis in 
meteorites (see below), it is now necessary to include many other species if
we wish to make a detailed comparison with real AGB stars. Calculations
including species beyond the CNO group are just becoming available now 
(but have been available for massive stars for quite some time), 
and although we will
deal with this in more detail below, the case of $^{26}$Al 
has been 
considered in the literature and is worthy of particular attention at 
this point.

The beta decay of $^{26}$Al produces 1.8 Mev $\gamma$-rays
(see Sch\"onfelder and Varendorff 1991). These
can be analysed
to determine the approximate amount of \Al26~present in the galaxy, with
current estimates giving $\sim 3-5 \msun$ (Clayton 
and Leising 1987). Since \Al26~has a half-life of $\tau_{26} \sim 10^6$ years,
this means there is about $2 \msun$ of \Al26~ejected into the Galaxy
every $\tau_{26}$ (Prantzos 1995). Many sources have
been postulated for this \Al26, and the analysis by Prantzos shows
that the distribution of \Al26~follows the spiral structure
of the galaxy, thus implying that it is associated with massive stars.
This is consistent with production by massive AGB stars as
well as Type~II supernovae and Wolf-Rayet stars.

Restricting our attention to AGB stars, there are two proven sites of formation
of \Al26;~these are the H-shell itself, and the bottom of the convective
envelope. The latter will be discussed below in the section on HBB, 
but it is important to note that the H-shell produces some
\Al26~via the Mg-Al cycle by transforming any initial $^{25}$Mg into
\Al26. This was investigated by Forestini {\it et al.\/} (1991), who
found that small amounts of \Al26~can be made and then dredged to 
the surface (although they had to force the dredge-up, which does not occur 
in their models). More recently, Gu\'elin {\it et al.\/} (1995) have observed
IRC+10216 for Mg and Al isotopes. They also present models of AGB stars
with HBB, and we defer a discussion of these models until Section 6.2

\section{Hot Bottom Burning in AGB Stars}

It has been known for some time that it was theoretically possible for
the convective envelope of a  star to reach so close to the H burning
shell that some nuclear
processing could occur at the bottom of the envelope. 
Cameron and Fowler (1971) suggested a mechanism for the production of
\Li7~which required HBB. In this picture, the \Hethree~left in the star
from earlier H-burning can capture an alpha particle at the base of the
convective envelope to form \Be7. If this \Be7~remains exposed
to high temperatures then it can capture a proton, and go on to complete
the PPIII sequence (see Figure 6). Alternatively, if the \Be7~decays into \Li7
then the \Li7 can capture a proton to complete the PPII sequence. If, however,
we are to make much \Li7~without completing the PP chains, then the 
\Be7~must be moved away from the hot region so that it can decay into \Li7. 
This \Li7~is also very fragile, and must spend most of its time in 
cool regions or it will be destroyed by the PPII chain. Clearly a 
convective envelope with a thin, hot base, can fulfill these criteria, 
and this is exactly what was proposed by Cameron and Fowler.
Indeed, there 
were some calculations carried out in the 70s by Sackmann {\it et al.\/} (1974) and
Scalo {\it et al.\/} (1975),
but with no observational motivation the models were not further studied
until recently. 

\begin{figure}
\epsfxsize = 0.7\hsize
\centerline{\quad\epsffile{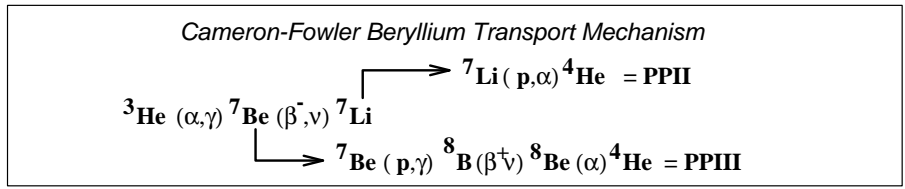}}
\smallskip
\caption{The Cameron-Fowler Beryllium Transport Mechanism.}
\end{figure}

\subsection{Observational Motivation}

Perhaps the first serious consideration of the possibility of HBB was
in the paper by Wood {\it et al.\/} (1983),  looked at the
very brightest AGB stars in the Magellanic Clouds and found that they were
not Carbon stars. In this
picture the brightest stars would have experienced many thermal pulses, 
and hence dredge-up episodes, as they ascended the AGB. So how could the 
brightest not have dredged enough \C12~to become Carbon stars?  
Wood {\it et al.\/}
suggested that HBB was responsible:~with a sufficiently hot 
envelope some CN cycling could occur, and conceivably process
the added \C12~into \N14 (predominantly).

Later, Smith and Lambert (1990) checked these stars for \Li7, 
an expected by-product of HBB via the Cameron-Fowler mechanism mentioned 
above. They found that all of these bright AGB M-stars 
showed extremely strong Li lines\footnote{The suggestion that perhaps 
these are supergiants rather than 
AGB stars is easily refuted, because they
show excesses of $s$-process elements (Smith and Lambert 1986) 
which  we have seen are 
also produced by thermal pulses on the AGB.}.
At about the same time there appeared some calculations which indicated
that the correct conditions did occur in some stars
(see, for example, Bl\"ocker and Sch\"onberner 1991; 
Lattanzio 1992). In models
of relatively large masses, above $\sim 5 \msun$, the convective envelope 
was seen to reach into the top of the H-burning shell, and hence the
material in the envelope was exposed to very high temperatures, reaching
up toward $T_6 = 100$! An example, for a $6 \msun$ model with $Z=0.02$ 
is shown in Figure~7. (This model will
be used throughout the rest of this paper to illustrate the various 
topics we discuss.)
Note that the temperature rises rapidly at first, as the
pulses reach ``full amplitude'', after which the growth is slowed somewhat. 
However, we see that even after 18 pulses the peak temperature during the 
interpulse phase is still growing, and is already above $T_6 = 80$.

\begin{figure}
\epsfysize = 0.4\vsize
\centerline{\quad\epsffile{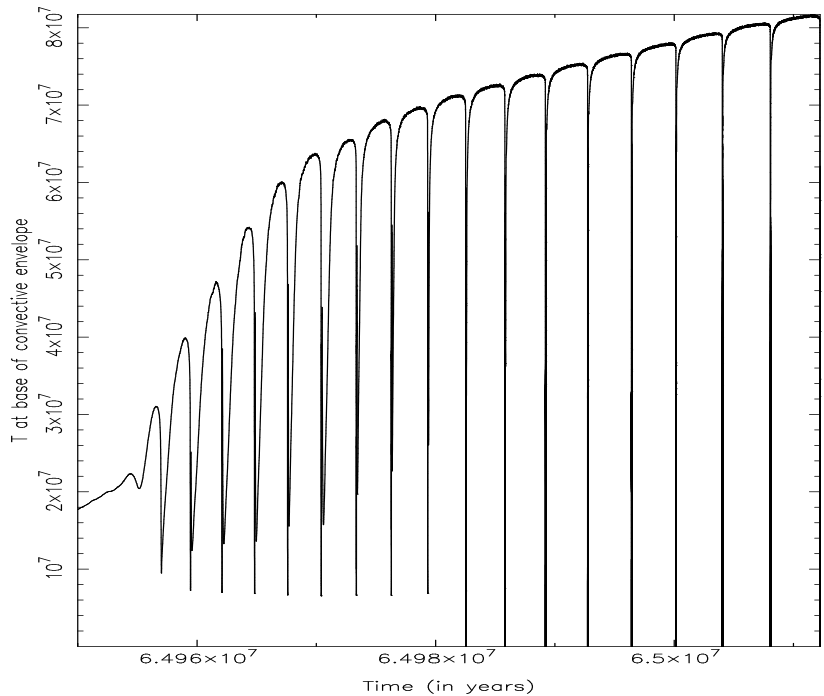}}
\smallskip
\caption{Temperature at the base of the convective envelope during 
the first few pulses of a $6\msun$ model with $Z=0.02$.}
\end{figure}

\subsection{The Production of Lithium}

Although there were early calculations of HBB and possible \Li7~production
(Sackmann {\it et al.\/} 1974, and
Scalo {\it et al.\/} 1975) the calculations of Boothroyd and 
Sackmann (1992a) showed 
quantitatively that such a scenario can work in the required stars. 
The peak abundances of \Li7~found by the models agreed very well with the 
observations, showing $\log \epsilon(^7{\rm Li}) \simeq 4.5$
\footnote{$\log\epsilon({\rm E}) = \log_{10} (n({\rm E})/n({\rm H)}) + 12$ 
where $n($E) is the number abundance of element E and $n($H) is the
number abundance of hydrogen.}.
After rapidly reaching the peak, however, the \Li7~is destroyed as it is 
repeatedly cycled through the hot bottom of the convective envelope. 
Also, the initial \Hethree~supply is finite, and once it is used there 
is no more to form the required \Be7. This behaviour is shown in Figure~8 
for the $6\msun$ model discussed above. These two effects combine to limit the 
lifetime of the so-called ``super-Li-rich giants'', so that they
only appear in a small range of $M_{bol} \simeq -6.2$ to $-6.8$.
This predicted range of luminosities
agrees well with  the observations for the Magellanic Clouds 
which showed Li-rich stars confined to a range
$M_{bol} \simeq -6$ to $-7$ (Smith and Lambert 1989, 1990).

A key ingredient in these calculations is the inclusion of some time-dependent
mixing algorithm. Boothroyd and Sackmann (and earlier authors) have used 
the diffusion equation to calculate the distribution of \Be7~and \Li7
in the convective regions. (They also quite nicely illustrated that
the instantaneous mixing assumption is incorrect for these species, 
and results in a decrease of \Li7 rather than an increase.)
In the calculations of Cannon {\it et al.\/} (1995) a
slightly different algorithm was used, which allows for different
compositions in the upward and downward moving streams, and some horizontal
diffusion at a given level.
This
reduces to the diffusion approximation in the case of infinitely quick 
horizontal diffusion between the two streams, but it does allow us to calculate
the different compositions in the upward stream, which has just been 
exposed to high temperatures, and the downward stream, which has 
been through the entire envelope and is now returning to the high 
temperature regions. The effect of this can be seen in Figure~9, where
the same algorithm as Cannon {\it et al.\/} is applied to the 
$6 \msun$ model with $Z=0.02$. 
The dashed lines represent the upward 
moving stream, which is richer in \Be7~(having been produced at the bottom
of the envelope) and poorer in \Li7~(which has just been destroyed at the 
bottom of the envelope by  the conclusion of the PPII chain). The solid line 
shows the downward stream, which is richer in \Li7~and poorer in \Be7~(due to the 
decay of the \Be7~into \Li7~in the outer, cooler parts of the envelope).

  \begin{figure}
\centerline{
     \hskip -1.0 truecm 
           \epsfxsize = 0.53\hsize\epsffile{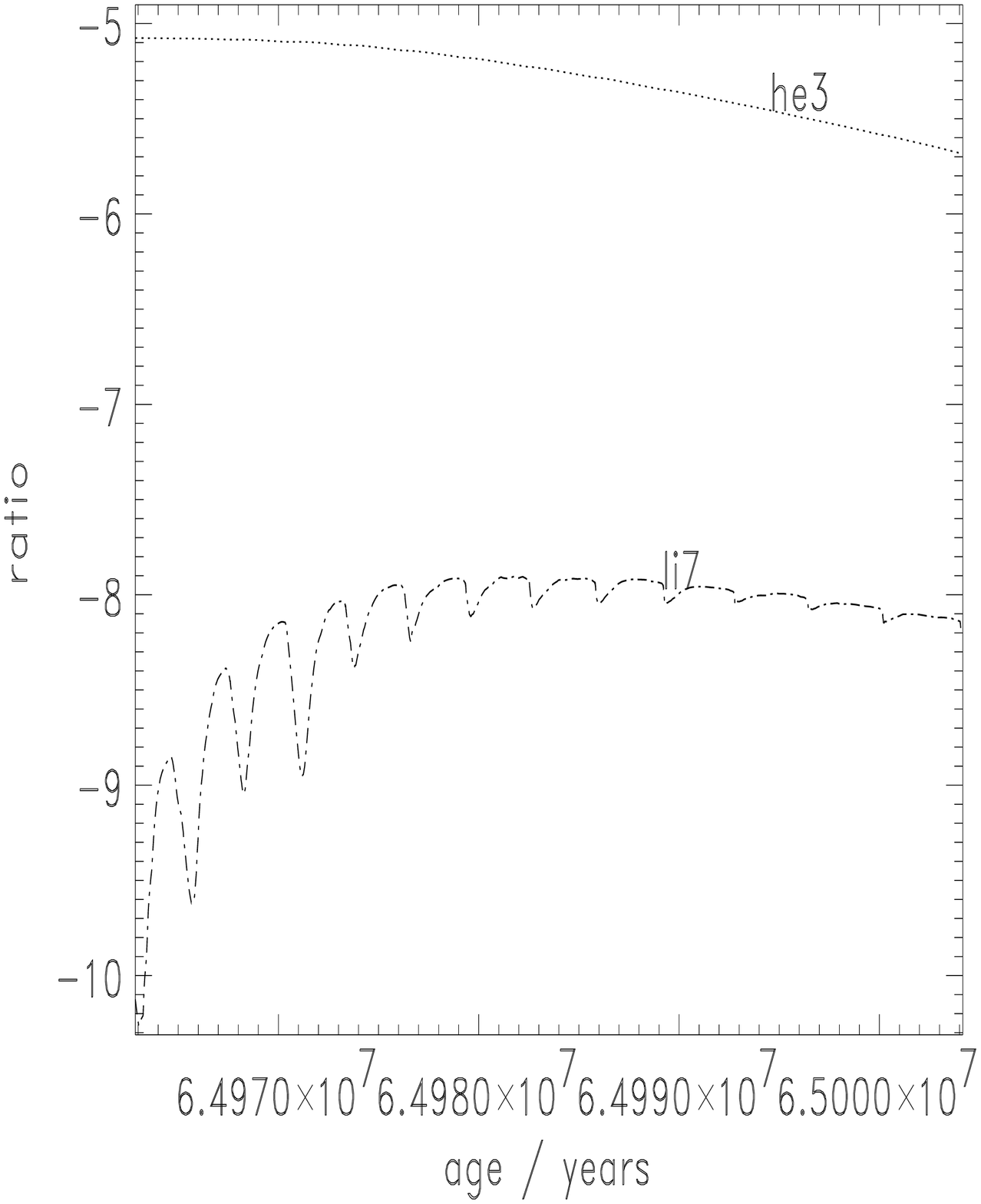}
     \hskip -0.1 truecm
           \epsfxsize = 0.53\hsize\epsffile{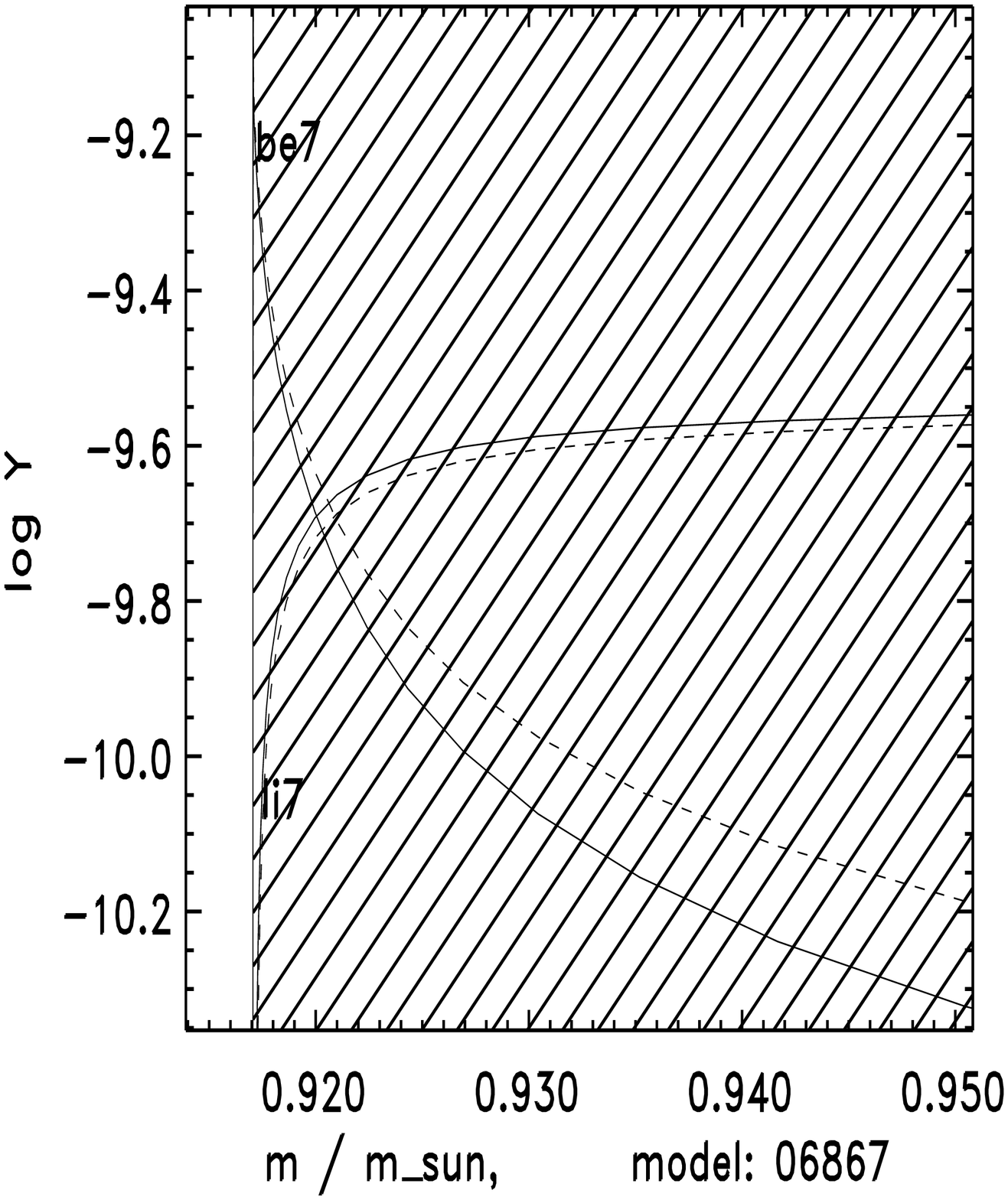}
}
  \begin{minipage}{8cm}
  \caption{Surface \Li7~and \Hethree~abundances in the
$6\msun$ model discussed in the text.}
  \end{minipage}
  \hfill
  \begin{minipage}{8cm}
  \caption{\Be7~and \Li7~abundances in the envelope of the $6\msun$  model 
discussed in the text. The hatched region denotes convection.}
  \end{minipage}
  \end{figure}

\subsection{Preventing Carbon Star Formation}

The original motivation for HBB by  Wood {\it et al.\/} (1983) was that it may
process sufficient \C12~for the formation of a Carbon star to be avoided,
despite the expected large amounts of \C12~dredged to the surface. This was
quantitatively shown to occur by Boothroyd,  {\it et al.\/}
(1993). They found
that HBB begun at $M \simeq 5\msun$ for $Z=0.001$. 
Models of $4\msun$ became Carbon stars
quite easily, but 
$5 \msun$ models
of this composition {\it just\/} failed to
become Carbon stars due to  the CN cycling at the bottom of the envelope. This
is
able to destroy all the \C12~added by each pulse. In fact, the envelope
quickly reaches the equilibrium value of \C12/\Cthirteen $\simeq 3$.
That is, the entire envelope is processed through the bottom of the
envelope sufficiently often during the interpulse phase that equilibrium 
abundances of \C12~and \Cthirteen~result. There is also, of course, some 
processing into \N14, but the key result is that HBB prevents the 
formation of Carbon stars for relatively massive AGB stars. 

The composition dependence of their results is very interesting.
For $Z=0.001$ there is negligible CN cycling by HBB at $4\msun$, which
indeed becomes a Carbon star.
However HBB in the $5\msun$ model prevents it  from becoming a 
Carbon star. At $6\msun$
the HBB is so efficient that from the first pulse the C/O ratio declines, 
and N/O rises dramatically, even exceeding unity. The behaviour at $Z=0.02$ is
similar. There is minimal dredge-up seen at $4\msun$, reflecting the
well-known phenomenon that dredge-up is more easily obtained at lower
metallicities. And where there is dredge-up, the increase in C/O is 
slower due to the higher \O16~abundance initially present. 
HBB begins at about $5 \msun$ again, but it mainly affects the 
\C12/\Cthirteen~abundance, with a negligible change in the C/O ratio.
In fact, the \C12/\Cthirteen~does not even reach equilibrium before 
mass-loss has removed the entire envelope. The $5\msun$ model
shows very substantial HBB, as did the same mass at $Z=0.001$.
Similar results are found for the $6\msun$ case discussed in this paper,
and the ratios of these surface abundances are shown in Figure~10.

Before leaving this subject, we mention briefly some problems associated 
with synthetic evolutionary calculations. In these, one parametrises the 
results of detailed evolutionary calculations ({\it e.g.\/} Iben 1981,
Renzini and Voli 1981, Groenewegen and de Jong 
1993, Marigo {\it et al.\/} 1995a,b) and then constructs stellar populations
for comparison with observations. 
Such calculations have led some to conclude that $M_c^{min}$, the minimum 
core-mass for dredge-up of \C12, is closer to $0.58\msun$ than the values
of $\sim 0.65\msun$ obtained in detailed evolutionary calculations.
Also, they conclude that the dredge-up parameter, 
$\lambda = {{\Delta M_{dredge}}/{\Delta M_H}}$ (see Figure~3b), is closer
to 0.6 than the value of 0.3 returned by evolutionary calculations.
We warn here against a literal interpretation
of these results: both $\lambda$ and $M_c^{min}$ are functions of composition,
mass, and mass-loss history (and mass-loss is not properly understood 
at present).
However, there is another problem associated with the input for these models.
At present we do not know when HBB ceases, or when dredge-up ceases.
Both of these require a reasonably massive envelope mass, but with almost
any combination of core-mass and envelope mass possible, due to differing
mass loss formulae, it is almost impossible to determine when these
effects cease. The predictive power of these models is weakened by
their dependence on which assumptions are made here.

\subsection{Core-Mass Luminosity Relation}

Finally, another consequence of HBB is that it results in departure from the 
well established core-mass---luminosity relation, which was discovered by 
Paczynski (1970a,b), and relates the maximum pre-flash surface luminosity to the
mass of the H-exhausted core. Although the original relation as quoted by 
Paczynski was independent of composition, numerous authors have refined 
his calculations and now the most accurate relations include the
effects of composition. It is the utility of this equation which stands at the
base of all the synthetic calculations, discussed above.

However, 
Bl\"ocker and Sch\"onberner (1991) showed
that once HBB begins the stars
no longer obey this relation. They followed a different
relation with a gradient at least 
a factor of 10 steeper. This were confirmed by Boothroyd and 
Sackmann (1992) and Lattanzio (1992). Bl\"ocker and Sch\"onberner found that
the reason for departure from the erstwhile  relation was that a very
deep convective envelope does not allow for a radiative zone which decouples 
the envelope from the H-shell. 
There are two obvious consequences of this discovery. 
Firstly, new synthetic evolutionary 
calculations will have to include this effect. Although the duration may be
short-lived, it can generate very high luminosities, and since 
mass-loss\footnote{We do not discuss the 
thorny issue of mass-loss in this paper.
For a discussion particularly relevant to AGB stars see Bl\"ocker (1995).} 
is tied to the luminosity, the mass loss increases also. Of course, as 
the envelope mass decreases, the HBB will eventually cease, and the model will
then return to the normal core-mass---luminosity relation. 
The second important consequence is that the previously assumed maximum 
luminosity on the AGB will be incorrect. It was assumed that once the 
core-mass reached the Chandrasekhar limit, then  the core would collapse and 
the star would leave the AGB as a supernova. So inserting this 
core-mass into the core-mass---luminosity relation yields a maximum 
luminosity for stars on the AGB. This is no longer correct, which means that
observational surveys may have {\it missed\/} the brightest AGB stars~!

We defer the discussion of HBB and oxygen ratios to Section 6.2.

\section{Constraints on Nucleosynthesis from Meteorite Grains}

In recent years we have seen the advent of a new source of information about 
the composition of stellar material.  This has been provided by measurements
of isotopic and elemental abundances in
in meteorites.
A significant advantage of these measurements is that they can provide
 information
about many elements for each grain, and since each grain has condensed in the
outflow from a single star, we obtain much compositional information from
a {\it single\/} stellar source. A further advantage is that  abundances 
can be found for species which are simply not visible in spectra. The 
disadvantage\footnote{There is always a disadvantage$\ldots$} is that 
we do not
know {\it a priori\/}
which kind of star produced which grain.
Although a young 
field (the first 
grain isolation occurred in 1987~! see Lewis {\it et al.\/} 1987) 
there is far too much
literature to be reviewed here. We will just try to give a flavour for the kinds of
constraints which the measurements place on the models, and then discuss some of the
recent models which attempt to address the problems. Further information is found in
Anders and Zinner (1993), Ott (1993) and the many papers in Section V of the 1994
{\sl Nuclei in the Cosmos III\/} meeting, edited by Busso, Gallino and Raiteri.

The grains of interest to us come from carbonaceous chondrites, and are
called ``exotic'' by meteoriticists, because of their isotopic anomalies 
compared to the solar system abundance distribution. These are the 
silicon carbide (SiC)
grains and the oxide grains, especially corundum (Al$_2$O$_3$). The other main
category, the graphite grains, are 
probably formed mainly in very
massive stars, as discussed by Travaglio and Gallino (1995).
Hence we do not discuss them here.

  \begin{figure}
\centerline{
     \hskip -1.7 truecm 
           \epsfysize = 0.40\vsize\epsffile{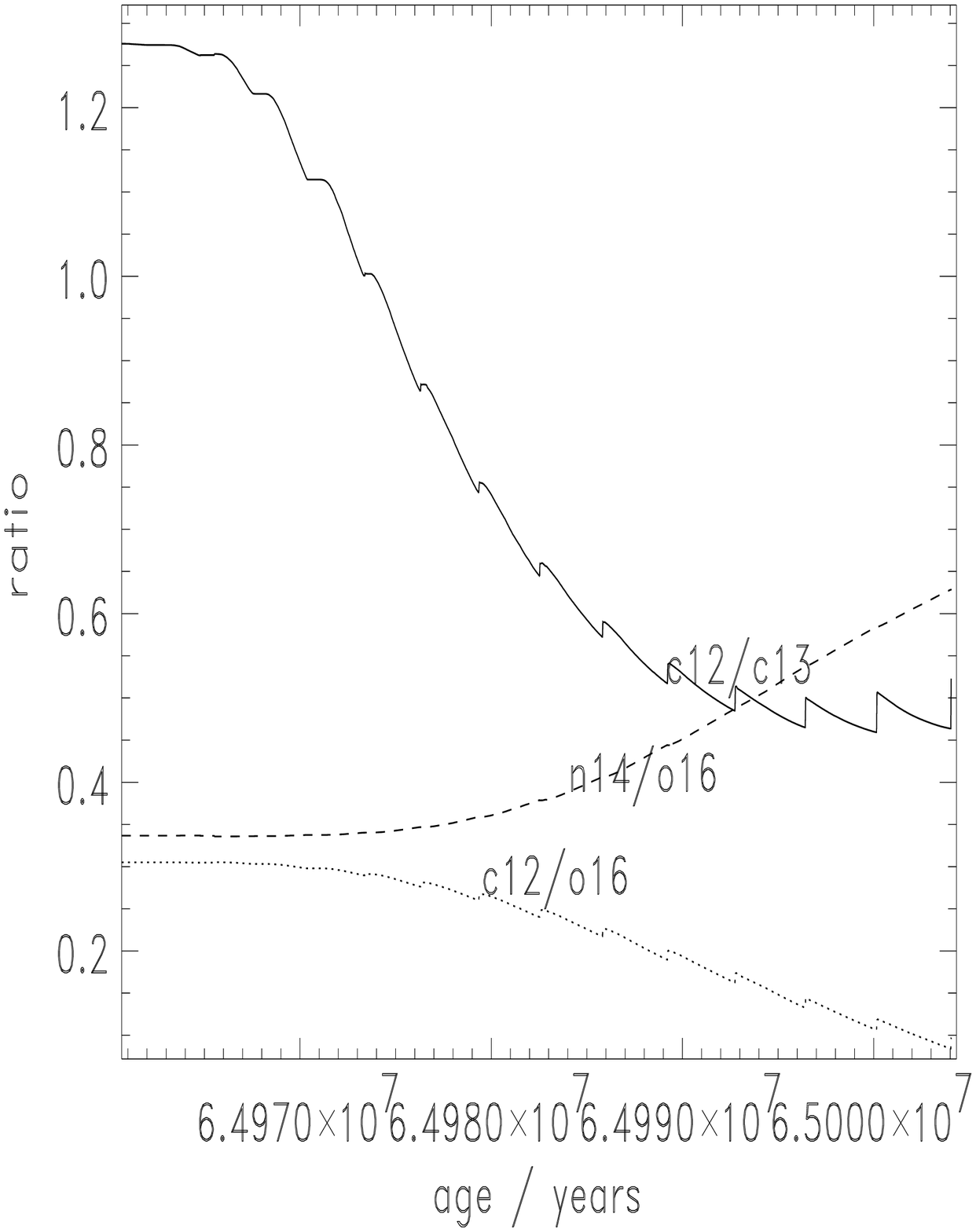}
     \hskip 1.4 truecm
           \epsfysize = 0.40\vsize\epsffile{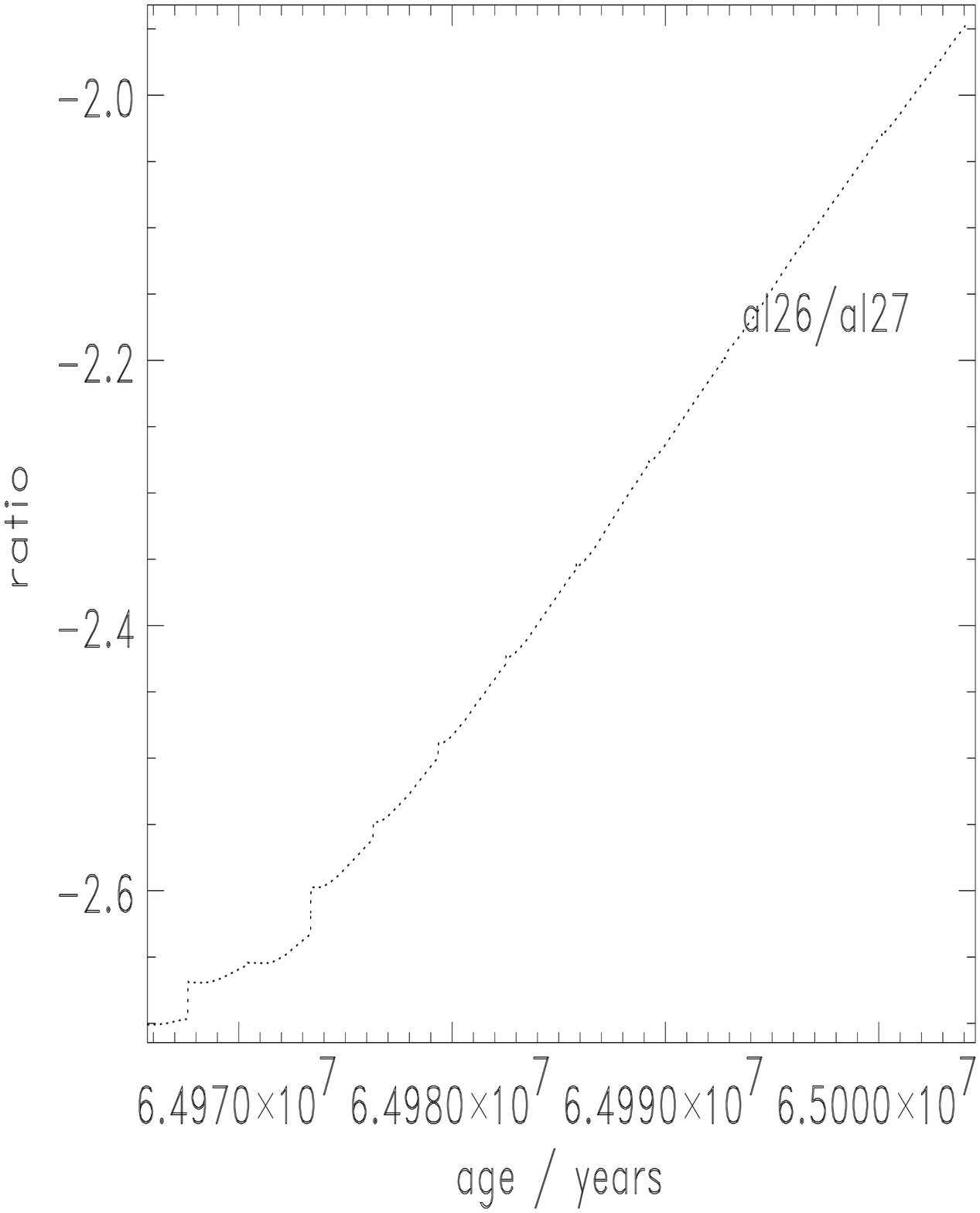}
}
  \begin{minipage}{8cm}
  \caption{Variation of surface ratios in the envelope of the $6\msun$
model discussed in the text. Note that the curve
labelled c12/c13 is actually log(\C12/\Cthirteen).}
  \end{minipage}
  \hfill
  \begin{minipage}{8cm}
  \caption{The increase of $\log($\Al26/$^{27}$Al) during the thermally 
pulsing evolution of the $6\msun$ model discussed in the text.}
  \end{minipage}
  \end{figure}

\subsection{SiC Grains:~Carbon and Nitrogen abundances}
Because the SiC grains must form
in a carbon-rich environment it is believed that these grains originated
in the envelopes
of Carbon stars. Further evidence comes from the distribution of 
\C12/\Cthirteen~ratios
in the grains, which is similar to that seen in Carbon stars, and the fact that
SiC is observed in the spectra of the dusty envelopes of many Carbon stars.

Basically, the abundances of the
carbon and nitrogen isotopes in these grains agree quite well
with predictions from stellar models. Nevertheless, there are some grains 
showing \C12/\Cthirteen~ratios less than the value expected from first 
dredge-up 
($\sim 20$), even going as low as 2 or 3, appropriate to CN equilibrium. 
Yet these
grains do not show \N14/\Nfifteen~ratios expected from CN burning:~they show 
ratios 
which are up to a factor of ten lower than expected from first dredge-up 
({\it e.g.\/} El~Eid 1994), and if
CN cycling is to reduce the \C12/\Cthirteen~ratios to the equilibrium values,
then these grains should be even richer in \N14. We have seen that HBB 
can produce 
\C12/\Cthirteen~ratios appropriate to equilibrium CN cycling, but these
models fail to account for the low \N14/\Nfifteen~ratios 
({\it e.g.\/} Sackmann and Boothroyd 1992, Boothroyd {\it et al.\/} 1994). 
Further, HBB is expected
to prevent the formation of Carbon stars by cycling the \C12~into 
\Cthirteen~and 
\N14, so how could SiC grains form in this environment? 
Is it possible that the J-stars could be the sources of these SiC grains with 
low \C12/\Cthirteen~ratios? And the $^{28,29,30}$Si isotopes
themselves are not seen in the ratios expected for neutron captures in the
intershell zone of thermally pulsing AGB stars,
and seem to indicate a spread of Si abundances in the initial composition.
Much more quantitative work needs to be done on stellar models to explain
all the data from these grains.

\subsection{Corundum Grains:~Oxygen and Aluminium abundances}
The most studied of the various oxide grain is corundum, Al$_2$O$_3$. 
The interest
in these is due to their oxygen and aluminium 
isotopic ratios, which show evidence of the three
dredge-up episodes, together with HBB. 

Nittler {\it et al.\/} (1994, 1995) divide these grains into 4 groups, according to their 
oxygen isotopic
ratios. The largest is Group~1, which shows \Oeighteen/\O16~and
 \Oseventeen/\O16~ratios
that are consistent with the first and second dredge-up:~namely that \Oeighteen/\O16
is reduced by a modest amount (less than a factor of two), while the 
\Oseventeen/\O16~ratio increases by a factor of up to twenty, compared to the
initial values of \Oeighteen/\O16$ \simeq 0.002$ and \Oseventeen/\O16 
$\simeq 0.00038$.
It appears that a satisfactory explanation
of all these grains requires us to consider stars of varying initial masses and 
a spread in the initial oxygen isotopic ratios, as discussed by 
Boothroyd {\it et al.\/} (1994).

The effect of HBB 
has been calculated by Boothroyd {\it et al.\/} (1995).
Initially there is a
rapid destruction of  the \Oeighteen~in the envelope, with a slower increase in 
the \Oseventeen~as the temperature at the base of the convective envelope 
increases with the  subsequent pulses. This would appear to explain 
many of the Group~2 grains, which have much lower \Oeighteen/\O16~ratios, 
and which simply
cannot result from
the first or second dredge-up episodes. Nevertheless, there are some Group~2 grains
that could only be explained by HBB if the lowest mass for HBB is substantially 
lower than is found in detailed models. For these, Boothroyd {\it et al.\/} (1995) suggested
that some deep mixing could be the explanation, a phenomenon they called ``cool
bottom burning'', and which is discussed below.
The Group~3 grains are a separate problem, showing  \Oseventeen/\O16~less than solar.
No satisfactory stellar site has been found for these grains yet, and they are
likely not from AGB stars.
The newly identified (Nittler {\it et al.\/} 1995) Group~4 grains show enhancements of
both \Oeighteen~and \Oseventeen. These could be due to AGB stars, as early thermal 
pulses can produce large amounts of \Oeighteen, and if the stellar mass is less than
about $5\msun$,  there is no HBB to destroy it. But the origin of these
grains is still unclear at present.

Corundum grains can be used for extracting information about heavier species as well.
For example, many of the oxide grains of Nittler {\it et al.\/} (1995) have been analysed for
Al-Mg and show excesses of $^{26}$Mg  but have normal  $^{25}$Mg/$^{24}$Mg. This
indicates that live \Al26~has decayed {\it in situ\/} to produce the  $^{26}$Mg.
The inferred initial \Al26/$^{27}$Al ratios are as large as 0.016. In Figure~11 we show
the variation in this ratio for our $6\msun$ model discussed earlier. The second
dredge-up raises the ratio from essentially zero to  $\sim 10^{-3}$, and once 
HBB begins the ratio climbs steadily, with  no sign of levelling off
when the calculations were terminated. It still shows
normal ratios of \Mg24/$^{25}$Mg, as required.
We should note that the largest values for 
\Al26/$^{27}$Al which can be obtained by dredging-up the products of the Mg-Al cycle 
from the H-shell are $\sim 2 \times 10^{-3}$ (Forestini {\it et al.\/} 1991). It appears that the
$6\msun$ model presented here would give rise to extreme Group~2 grains, with
the current values of \O16/\Oeighteen~$\sim 10^6$ and 
\O16/\Oseventeen~$\simeq 100$. Note that these values agree well with the
$5\msun$ models of Gu\'elin {\it et al.\/} (1995), as well as their
observations of IRC+10216.

There is also a wealth of data available for other species, such as Ti, Xe, etc
(see, for example, Gallino {\it et al.\/} 1994)
but this will not be addressed here, except to remind us that we ignore this
new source
of highly accurate and wonderful data at our peril.

\subsection{Cool Bottom Burning}

We mentioned above that some of the Group~2 corundum grains seem to imply that
HBB occurs in masses which are too small to be consistent with the extant models.
A possible solution to this was proposed by Boothroyd {\it et al.\/} (1995) 
and followed up
by Wasserburg {\it et al.\/} (1995). In this model there is some extra 
mixing from
the bottom of the convective envelope down toward the H-shell. This is
called ``Cool Bottom Burning'', and seems to produce the
low values of  \Oeighteen/\O16~required by the grains (and also many observations
of Carbon stars, which also show similar isotopic ratios). Wasserburg {\it et al.\/} (1995)
showed that the isotopic ratios 
depend critically on the temperature (of course!) to which the material is mixed,
but there was very little dependence on the rate of mixing. For best results
the mixing reaches down to $\Delta \log T \simeq 0.17$ from the base of the H-shell
on the AGB. 

Note that Charbonnel (1994) has 
suggested a similar mechanism to explain the anomalously
low \C12/\Cthirteen~ratios for low mass  stars, when compared to the predictions 
from first dredge-up calculations. Wasserburg {\it et al.\/} (1995) 
found that an identical
mixing on the first giant branch, with the same $\Delta \log T$, produced
\C12/\Cthirteen~ratios in the required range. Further studies are
needed.

\section{The Future: What Should Be Done ?}

So, after that lengthy introduction, we come to the main topic of this review!
We will break this into two subsections.

\subsection{Evolution}

There remain many uncertainties in modeling of AGB stars. First and foremost is
the lack of a good theory of convection (still). For AGB stars the thing which
we most need is an accurate way to determine the boundaries of the various
convective zones, and any associated overshoot. Note that changes in the
assumptions
one uses to treat convection can make large differences in the amount of
dredge-up obtained (Frost 1995). Various authors may use different 
ways of treating a discontinuity in $\nabla_{rad}/\nabla_{ad}$ at the edge of a 
convective zone. These may all be physically motivated, and phenomenologically 
reasonable, yet yield different results. 

Sometimes during dredge-up 
we can obtain convergence problems. If one iterates on the physical variables
until converged, and then mixes in the convective zone, one has a model
which is internally inconsistent:~the composition used for convergence is not
that which resulted from the implied mixing. Alternatively, one could
alternate iterations on the physical variables with mixing over the current
convective zone. (This is what we do.) When this works, one has an internally
consistent model. Yet sometimes this procedure will not converge, and some
other recourse must be taken and, to some extent,  the amount of 
dredge-up obtained depends on how these problems are handled (see Frost 1995).
Of course, there are various other schemes,
such as mixing first and then iterating (Sweigart 1995). In short, these
differences in treatment of details in the convection, as well as treatment
of convective boundaries, can explain the differences in the size of the
dredge-up parameter found by various authors.

Probably related to this is the expected (but rarely seen) semiconvective
mixing of~H~down beyond
the formal convective boundary during dredge-up. It is this which produces the 
(apparently) required \Cthirteen~pocket responsible for the neutrons
that enable the $s$-processing to occur on the AGB. We need to know how this
pocket is formed, and its size. All of this seems to require a greater knowledge
of convection and mixing (again, especially at boundaries, and probably
involving semiconvection) than we have at present.

Another convection  problem is that we need to know how 
the dredge-up varies with mass, composition
and mass-loss history. Yet, as outlined above, we cannot even agree on
how to calculate it, let alone embark on a computing job of 
such magnitude (Renzini 1989).

\subsection{Nucleosynthesis}

With such a wealth of  data now available, both from stars and meteorite grains, 
it has become clear that for quantitative comparisons we must use codes which
contain many more nuclear species than are usually included in evolutionary
calculations. It is now relatively common to see calculations now including
20 to 40 species, and our nucleosynthesis calculations reported in this paper
for a $6 \msun$ model use a network of 74 species and some 506 reactions.
It seems that calculations of this size are now the minimum we need for
comparison with the avalanche of data coming our way.

Of course, all calculations of nucleosynthesis are only as accurate
as the rates used. Some of the most important rates for these
calculations ({\it e.g.\/} the Ne-Na and Mg-Al cycles) are not very well known yet,
as reviewed by Arnould (1995). These data are required urgently.

The consequences of the radiative burning of \Cthirteen~need to be
investigated. How does this affect the $s$-processing? How
will it affect other nucleosynthesis? Indeed, how does the \Cthirteen~pocket
come into existence? 

It also appears that we can no longer assume that all mixing occurs
instantaneously. We have seen that this 
assumption must be removed to produce the Li-rich 
stars. A similar situation is likely to exist for some species in the intershell
convective zone. Although this time-dependent mixing is likely to have
little effect on the stellar structure (as most of the reactions involved
are energetically negligible), it may well be crucial for accurate calculations
of the nucleosynthesis. In lieu of a suitable theory of time-dependent convective
mixing, we must use the diffusion equation ({\it e.g.\/} Boothroyd {\it et al.\/} 1995) or some
variant (Cannon {\it et al.\/} 1995, Wasserburg {\it et al.\/} 1995).

Related to this is the postulated ``cool bottom burning'', where material 
burns while moving (slowly) through radiative zones. This should be investigated
in two ways:~firstly phenomenologically, to see if it can account for the
observed abundances, and secondly
from a purely physical view, so that the mechanism which drives
the mixing can be understood!

\section{Conclusion}

We hope we have conveyed some of the excitement, and frustration, of AGB stellar 
modeling. We are on the verge of a new level of quantitative understanding, 
produced by accurate observations and meteoritic measurements, which have spurred 
the theorists to include more and more species and even try to calculate time-dependent 
mixing. Almost all of the areas we listed for future research are indeed being
investigated by various workers at present, but the most serious problem remains
that of determining convective boundaries under the complex conditions found 
in these stars. The extant hydrodynamical studies of convection
({\it e.g.\/} Nordlund 1995, Zahn 1995) cannot cover the
dynamical range of densities required, nor match the other conditions
(such as viscosities) seen in the stellar context.
Sadly, we see no reason for optimism in this area in the near future.

But we cannot end on a sad note. We have made enormous strides in the last few years,
and we now have a new source of information concerning abundances. The next few
years should be exciting.

%
%
\section*{Acknowledgments}
This research was supported in part by the Australian Research 
Council. CAF wishes to acknowledge the assistance of
an Australian Post-Graduate Award, and the British Council for travel funds to 
visit the Institute of Astronomy, Cambridge, where this work was initiated.
%
%
 
\beginrefer

\refer Anders, E., and Zinner, E.: 1993, {\sl Meteoritics\/}, {\bf 28}, 490.

\refer Arnould, M.: 1995, this volume.

\refer Bl\"ocker, T.: 1995, {\sl A. \& A.\/}, in press.

\refer Bl\"ocker, T., and Sch\"onberner, D.: 1991, {\sl A. \& A.\/}, 
{\bf 244}, L43.

\refer  Boothroyd, A. I. and Sackmann, I.-J.: 1992, {\sl Ap. J. Lett.\/}, 
{\bf 393}, L21.

\refer Boothroyd, A. I. {\it et al.\/}: 1993, {\sl Ap. J.\/}, {\bf 416}, 762.

\refer Boothroyd, A. I. {\it et al.\/}: 1994, {\sl Ap. J. Lett.\/}, {\bf 430}, L77.

\refer Boothroyd, A. I. {\it et al.\/}: 1995, {\sl Ap. J. Lett.\/}, {\bf 442}, L21.

\refer Cameron, A. G. W., and Fowler, W. A.: 1971, {\sl Ap. J.\/}, 
{\bf 164}, 111.

\refer Cannon, R. C., {\it et al.\/}: 1995,~in {\sl Nuclei in the Cosmos III}, 
Eds.  M. Busso, R. Gallino, C~.M. Raiteri, (AIP: New York), p.~469.

\refer Castellani, V., {\it et al.\/}: 1971a, 
{\sl Astr. Sp. Sci\/.}, {\bf 10}, 340.

\refer Castellani, V., {\it et al.\/}: 1971b, 
{\sl Astr. Sp. Sci.\/}, {\bf 10}, 355.

\refer Charbonnel, C.: 1994, {\sl A. \& A\/.}, {\bf 282}, 811.

\refer Clayton, D. D., and Leising, M.: 1987, {\sl Phys. Rep.\/}, {\bf 144}, 1.

\refer Deupree, R. G.: 1984, {\sl Ap. J.\/}, {\bf 287}, 268.

\refer El Eid, M.: 1994, {\sl A. \& A.\/}, {\bf 285}, 915.

\refer Forestini, M., {\it et al.\/}: 1991, {\sl A. \& A.}, {\bf 252}, 597.

\refer Forestini, M., {\it et al.\/}: 1992, {\sl A. \& A.}, {\bf 261}, 157.

\refer Frost, C. A.: 1995, Monash University Internal Report.

\refer Gallino, R.,  and Arlandini, C.: 1995, this volume.

\refer Gallino, R., {\it et al.\/}: 1988, {\sl Ap. J. Lett.\/}, {\bf 334}, L45.

\refer Gallino, R., {\it et al.\/}: 1994, {\sl Ap. J. \/}, {\bf 430 },858 .

\refer Groenewegen, M. A. T., and de Jong, T,: 1993, {\sl A. \& A.}, 
{\bf 267}, 410. 

\refer Gu\'elin, M., {\it et al.\/}: 1995, preprint.

\refer Iben, I., Jr.: 1975a, {\sl Ap. J.\/}, {\bf 196}, 525.

\refer Iben, I., Jr.: 1975b, {\sl Ap. J.\/}, {\bf 196}, 549.

\refer Iben, I., Jr.: 1981, {\sl Ap. J.\/}, {\bf 246}, 278.

\refer Iben, I., Jr.: 1982, {\sl Ap. J.\/}, {\bf 260}, 821.

\refer Iben, I., Jr.: 1991, in {\sl Evolution of Stars: The Photospheric 
Abundance Connection\/}, p.~257.

\refer Iben, I., Jr., and Renzini, A.: 1982a, {\sl Ap.~J.\/}, 
{\bf 259}, L79.

\refer Iben, I., Jr., and Renzini, A.: 1982b, {\sl Ap.~J.\/}, 
{\bf 259}, L79.

\refer Iben, I., Jr., and Renzini, A.: 1983, {\sl A. R. A. \& A.\/}, 
{\bf 21}, 271.

\refer Jorissen, A, {\it et al.\/}: 1992, {\sl A. \& A.}, 
{\bf 261}, 164.

\refer Lattanzio, J. C.: 1984, {\sl MNRAS\/}, {\bf 207}, 309.

\refer Lattanzio, J. C.: 1989, in {\sl Evolution of Peculiar Red Giants\/}, 
Eds. H. R. Johnson and B. Zuckerman, CUP, p.~161.

\refer Lattanzio, J. C.: 1992, {\sl Proc. Astron. Soc. Aust.\/}, 
{\bf 10}, No. 2., 120.

\refer Lauterborn, D., {\it et al.\/}: 1971, {\sl A. \& A.\/}, 
{\bf 10}, 97.

\refer Lewis, R. S., {\it et al.\/}: 1987, {\sl Nature\/}, {\bf 326}, 160.

\refer Marigo, P., and Chiosi, C.: 1995a, this volume.

\refer Marigo, P., and Chiosi, C.: 1995b, {\sl A. \& A. \/}, submitted.

\refer Mowlavi, N., {\it et al.\/}: 1995, this volume.

\refer Nittler, L. R., {\it et al.\/}: 1994, {\sl Nature\/}, {\bf 370}, 443.

\refer Nittler, L. R., {\it et al.\/}: 1995, in {\sl Nuclei
in the Cosmos III\/},
Eds. M. Busso, R. Gallino, C. Raiteri, AIP, p.~585.

\refer Nordlund, A.: 1995, this volume.

\refer Ott, U: 1993, {\sl Nature\/}, {\bf 364}, 25.

\refer Paczy\'nski, B.: 1970a, {\sl Acta Astron.\/}, {\bf 20}, 47.

\refer Paczy\'nski, B.: 1970b, {\sl Acta Astron.\/}, {\bf 20}, 287.

\refer Prantzos, N.: 1995, in {\sl Nuclei in the Cosmos III}, 
Eds.  M. Busso,
R. Gallino, C~.M. Raiteri, (AIP: New York), p.~553.

\refer Renzini, A.: 1989, in {\sl Evolution of Peculiar Red Giants\/}, 
Eds. H. R. Johnson and B. Zuckerman, CUP, p 413ff.

\refer Renzini, A., and Voli, M.: 1981, {\sl A. \& A.\/}, {\bf 94}, 175.

\refer Sackmann, I.-J.: 1977, {\sl Ap. J.\/}, {\bf 212}, 159.

\refer Sackmann, I.-J.: 1980, {\sl Ap. J. Lett.\/}, {\bf 241}, L37.

\refer Sackmann, I.-J., and Boothroyd, A. I.: 1991, in 
{\sl Evolution of Stars: The Photospheric  Abundance Connection\/}, p.~275

\refer Sackmann, I.-J., and Boothroyd, A. I.: 1992, {\sl Ap. J. Lett.\/}, {\bf } .

\refer Sackmann, I.-J., {\it et al.\/}:~1974, {\sl Ap. J.\/}, {\bf 187}, 555.

\refer Scalo, J. M., {\it et al.\/}: 1975,{\sl Ap. J.\/}, {\bf 196}, 809.

\refer Sch\"oenfelder, V., and Varendorff, M.: 1991, in {\sl Gamma-Ray Line
Astrophysics\/}, Eds. Ph. Durouchoux and N. Prantzos.

\refer Schwarzschild, M., and H\"arm, R.: 1965,{\sl Ap. J.\/}, {\bf 142}, 855.

\refer Smith, V. V., and Lambert, D. L.: 1986, {\sl Ap. J. \/}, 
{\bf 311}, 843.

\refer Smith, V. V., and Lambert, D. L.: 1989, {\sl Ap. J.\/}, {\bf 311}, 843.

\refer Smith, V. V., and Lambert, D. L.: 1990, {\sl Ap. J. Lett.\/}, 
{\bf 361}, L69.

\refer Smith, V. V., {\it et al.\/}: 1987, {\sl Ap. J.\/},
{\bf 320}, 862.

\refer Straniero, {\it et al.\/}: 1995a, in {\sl Nuclei in the Cosmos III}, 
Eds.  M. Busso,
R. Gallino, C~.M. Raiteri, (AIP: New York), p.~407.

\refer Straniero, O. {\it et al.\/}: 1995b, {\sl Ap.~J.~Lett.}, {\bf 440}, L85.

\refer Sugimoto, A., and Fujimoto, M. Y.: 1978, {\sl Pub. 
Astron. Soc. Japan\/}, {\bf 30}, 467.

\refer Sweigart, A. V.: 1995, personal communication.

\refer Travaglio, C.,  and Gallino, R.: 1995, this volume.

\refer Truran, J. W., and Iben, I., Jr.: 1977, {\sl Ap. J.\/}, {\bf 216}, 797.

\refer Wasserburg, G. J., {\it et al.\/}: 1995, {\sl Ap. J. Lett.\/}, {\bf 447}, 37L.

\refer Wood, P. R., {\it et al.\/}: 1983, {\sl Ap. J.\/}, 
{\bf 272}, 99.

\refer Zahn, J. P.: 1995, this volume.

\endrefer
\end{document}